\documentclass[review,5p]{elsarticle}
\usepackage{hyperref}
\usepackage{amsmath}
\usepackage{mathrsfs}
\usepackage{xfrac}
\usepackage{graphicx}
\usepackage{amsfonts}
\usepackage{bm}
\usepackage{xspace}
\usepackage{float}

\newcommand{\nm}{n}

\newcommand{\lat}{a_{\text{lat}}}

\newcommand{\startsquarepar}{%
    \par\begingroup \parfillskip 0pt \relax}
\newcommand{\stopsquarepar}{%
    \par\endgroup}

\begin{document}

\begin{frontmatter}

\title{Three-dimensional Gross-Pitaevskii solitary waves in optical lattices: stabilization using the artificial quartic kinetic energy induced by lattice shaking}

\author[umb]{M.\ Olshanii\corref{cor1}}
\ead{maxim.olchanyi@umb.edu}
\author[umb]{S.\ Choi}
\author[umb]{V.\ Dunjko}
\author[nu]{A.\ E.\ Feiguin}
%
\author[sorb]{H. Perrin}
\author[umb]{J.\ Ruhl}
\author[jpl]{D.\ Aveline}
\address[umb]{Department of Physics, University of Massachusetts Boston, Boston Massachusetts 02125, USA}
\address[nu]{Department of Physics, Northeastern University, Boston, Massachusetts 02115, USA}
\address[sorb]{Laboratoire de physique des lasers, CNRS, Universit\'{e} Paris 13,
Sorbonne Paris Cit\'{e}, 99 avenue J.-B. Cl\'{e}ment, F-93430 Villetaneuse, France}
\address[jpl]{Jet Propulsion Laboratory, California Institute of Technology, Pasadena, CA 91109, USA}
\cortext[cor1]{Corresponding author}

\date{\today}

\begin{abstract}
In this Letter, we show that a three-dimensional Bose-Einstein solitary wave can become stable if the dispersion law is changed from
quadratic to quartic. We suggest a way to realize the quartic dispersion, using shaken optical lattices. Estimates show that the resulting
solitary waves can occupy as little as $\sim 1/20$-th of the Brillouin zone in each of the three directions and contain as many as $N = 10^{3}$ atoms,
thus representing a \textit{fully mobile} macroscopic three-dimensional object.
\end{abstract}
\end{frontmatter}

\section{Introduction}
Creating mobile self-supporting three-dimensional (3D) matter waves---3D analogs of the 1D solitons\footnote{A \textit{solitary wave} is an isolated wave that maintains its shape due to a balancing of dispersion and nonlinear attraction; in strict usage \cite{ablowitz1981}, a \textit{soliton} is a solitary wave with a special property: when it collides with another local disturbance---e.g. another soliton---then asymptotically far from the collision, it regains its initial shape and velocity. Regardless of how they may be realized, 3D solitary waves are not expected to be solitons, because they are not expected to instantiate any integrable systems.} first realized in Refs.~\cite{strecker2002_150,khaykovich2002_1290}---is a long-standing goal of physics of ultracold gases \cite{ahufinger2004_053604,alexander2006_040401,saito2004_053610,mihalache2005_021601,borovkova2011_035602}.
3D Bose-Einstein solitary waves in continuum space are unstable because they violate \cite{mihalache2005_021601} the Vakhitov-Kolokolov (VK) stability criterion \cite{vakhitov1973_1020}. One strategy for circumventing this problem is to create discrete solitary waves in optical lattices. These objects may be stable for some sets of parameters \cite{ahufinger2004_053604,mihalache2005_021601,alexander2006_040401},
but are localized on a limited number of sites. Thus they occupy a substantial portion of the Brillouin zone and so have limited mobility. Another stabilizing strategy is to oscillate in time the interatomic coupling constant around a negative value, but switching to positive for periods of time \cite{saito2004_053610} (the VK criterion does not apply to time-dependent nonlinearities). However, here substantial atomic losses will limit the lifetime of the 3D object. In a third approach, 3D solitary waves are stabilized by varying the nonlinearity strength in space \cite{borovkova2011_035602}; but then they are not free objects moving in a translationally invariant medium.

Here we propose a way to satisfy the VK criterion by changing the atomic dispersion law from quadratic to quartic. We suggest using shaking to couple the lowest and the first-excited energy bands in a 3D optical lattice so that the quadratic portions of the dispersions cancel out, and the quartic terms become dominant. This results in a highly mobile localized object.

\section{Dispersion law and the stability of solitary matter waves}
Let us discuss how the dispersion law enters the VK criterion. Our main conclusion, given in Eq.~(\ref{VT_for_GPE}) below, may also be reached using the well-known heuristic argument based on how the kinetic and interaction energies scale with the typical length (we will summarize this reasoning following that equation); but because of the novelty of the situation, we prefer to use the more rigorous route via the VK criterion.
Consider a time-dependent nonlinear Schr\"{o}dinger equation for the wavefunction of a $d$-dimensional
Bose-Einstein condensate with a generalized kinetic energy represented
by a particular differential operator of $p$-th ($p$ being even) degree (below we will propose a  scheme, using ultracold gases, for realizing a physical system described by Eq.~(\ref{time-dependent_GPE}) with $d=3$ and $p=4$):
\begin{equation}
%
i \hbar \frac{\partial \Psi}{\partial t}    = \eta \sum_{i=1}^{d} (\hat{p}_{i})^{p}\Psi+g|\Psi|^2\Psi
\label{time-dependent_GPE}
\,,
%
\end{equation}
where $\eta>0$ is the dispersion strength, $g<0$ is the coupling constant,
and $\hat{p}_{i} \equiv -i\hbar\frac{\partial}{\partial r_{i}}$ is the operator of the $i$th component of momentum. Assume that Eq.~(\ref{time-dependent_GPE})
admits normalizable stationary solutions
\begin{align}
\Psi({\bm r},\,t) = \exp[-i \mu t/\hbar] \psi({\bm r})
\,\,:
\label{Psi_vs_psi}
\end{align}
their spatial part
$\psi({\bm r})$ will then be governed by a time-independent nonlinear Schr\"{o}dinger equation,
\begin{equation}
\begin{split}
& \eta \sum_{i=1}^{d} (\hat{p}_{i})^{p}\psi+g|\psi|^2\psi = \mu \psi
\label{GPE}
\\
& \eta>0,\,\,g<0. 
\end{split}
\end{equation}
Any solution of (\ref{GPE}) is a point of extremum, subject to the normalization constraint that
\begin{align}
\int \!d^{d}  {\bm r} \,\, |\psi|^2 = N
\label{N}
\end{align}
be the number of particles,
of the energy functional
\begin{align}
&
E = T + V
\,.
\label{E}
\end{align}
Here
\begin{align}
&
T = \eta \hbar^{p} \sum_{i=1}^{d}  \int \!d^{d}{\bm r} \,\,  \left|\frac{\partial^{p/2} \psi}{\partial r_{i}^{p/2}}\right|^2\quad\text{and}
\label{T}
\\
&
V = \frac{1}{2}g \int \!d^{d}{\bm r} \,\, |\psi|^4
\label{V}
\end{align}
are the kinetic energy and the interatomic interaction energy functionals, respectively. The corresponding variational space
is the space of continuous functions of the coordinates; the chemical potential $\mu$ enters as the Lagrange multiplier enforcing the constraint in Eq.~(\ref{N}).
We can conjecture a stability criterion for localized stationary solutions of Eq.~(\ref{time-dependent_GPE}) already at the level of the virial theorem: regarding the interaction functional in Eq.~(\ref{V}) as a mean-field approximation of the pairwise interaction energy
with the $d$-dimensional $\delta$-function as the potential, the virial theorem \cite{book_landau_mechanics} would predict $p T = -d V$,
and subsequently
\begin{align}
E = \frac{d-p}{d}\, T
\,\,.
\label{E_vs_T_virial}
\end{align}
Now, one would expect the energy of a stationary state to be
below the energy of a dilute, heavily delocalized cloud; the energy of the latter is close to zero,
so the right-hand-side of Eq.~(\ref{E_vs_T_virial}) must be negative.
At this point we conjecture that for solitary waves to exist, the dispersion law
must be sufficiently sharp, namely $p > d$. Next, we will confirm this using the rigorous VK criterion.

First, observe that if one multiplies the stationary nonlinear wave equation in Eq.~(\ref{GPE}) from the left by $\Psi^{*}$, integrates it over all space, and
combines the result with the expression for the energy given in Eqs.~(\ref{E},\,\ref{T},\,\ref{V}), one obtains   \cite{mihalache2005_021601} $E = \mu N - V$.
Combining this with Eq.~(\ref{E_vs_T_virial}), we obtain a relationship between the chemical
potential and the kinetic energy that, in particular, provides information about the sign of the chemical potential:
\begin{align}
\mu = \frac{d-2p}{d}\, \frac{T}{N}
\,\,.
\label{mu_vs_T}
\end{align}

The VK criterion deals with continuous families of stationary localized states parametrized by their norm.
The scaling properties of the constituents of Eq.~(\ref{GPE}) imply that members of such a family will be connected by scaling transformations. Indeed,
it is easy to show that if there exists a stationary solution $\psi_{1}({\bm r})$---corresponding to the chemical potential $\mu_{1}$ and the number of atoms $N_{1}$---of the
stationary wave equation in Eq.~(\ref{GPE}), then
\begin{align}
&
\psi({\bm r}) = \lambda(N)^{-p/2}\psi_{1}({\bm r}/\lambda(N)) \label{lambda_family1}
\\
&
\mu = \lambda(N)^{-p}  \mu_{1}
\,\,,
\label{lambda_family2}
\end{align}
with
\begin{align}
\lambda(N) = \left(\frac{N}{N_{1}}\right)^{\frac{1}{d-p}}
\,\,,
\label{lambda_vs_N}
\end{align}
is also a solution. According to the VK criterion, a member of the family of localized stationary solutions in Eqs.~(\ref{lambda_family1},\,\ref{lambda_family2})
is dynamically stable---in particular against collapse or dispersion to infinity---if its chemical potential decreases with the number of particles,
\begin{align}
\frac{d \mu}{d N} < 0
\,\,,
\label{VK}
\end{align}
where the derivative is taken along the family. Using the results in Eqs.~(\ref{mu_vs_T}
-\ref{lambda_vs_N}), it is easy to show that in our case,
\begin{align*}
\frac{d \mu}{d N}  =
-
\frac{p  (2p-d)}{d  (p-d)} \,\,  \frac{T_{1}}{N_{1}^2}  \,\, \left(\frac{N}{N_{1}}\right)^{\frac{d}{p-d}}
\,\,.
\end{align*}
So Eq.~(\ref{time-dependent_GPE}) supports stable stationary solitary waves if
\begin{align}
p > d
\mbox{\,\,\,     or\,\,\,     }
0 < p < \frac{d}{2}
\,\,.
\label{VT_for_GPE}
\end{align}
The second case does not currently seem physical. Recall that $p$ is required to be even; then, even for $d=3$, there is no $p$ yielding the second inequality.

The heuristic argument for the same conclusion goes as follows: let $\ell$ be the typical length scale of the atomic cloud. On dimensional grounds, the kinetic energy scales as $\sim +1/\ell^{p}$, while the interaction energy, which is proportional to the density, scales as $\sim -1/\ell^{d}$. The matter wave will be stable if the total energy has a minimum. 

Consider first the case $p>d$. Then the kinetic energy dominates for small $\ell$ ($E \to +\infty$), and the interaction energy for large $\ell$ ($E \to 0$). Consider the energy as a function of $\ell$, and assume, as is reasonable, that it is a smooth function of it. Then $dE/d\ell$ is negative for small $\ell$, and positive for large $\ell$; thus, for at least one value of $\ell$, $dE/d\ell$ switches the sign from negative to positive, so this value of $\ell$ is a local minimum. $dE/d\ell$ could cross zero at multiple values of $\ell$, but at the smallest and the largest of these, $dE/d\ell$ must cross from negative to positive values, in order to match the asymptotic behavior. In any case, there are local minima; moreover, because $E(\ell)$ is smooth, it follows that at least one of these local minima must be a global minimum.

Now assume $p<d$. To show that $E(\ell)$ can have no global or local minimum, additional assumptions will be needed: the reason is that while  $dE/d\ell$ must in this case switch sign from positive to negative at least once (and if more than once, the first and the last switch must be from positive to negative, to match the asymptotic behavior), there could, conceivably, also exist intermediate zeros of $dE/d\ell$ where the switch is from negative to positive. These would be local minima, which would mean a potentially stable matter wave. However, heuristically, it seems unlikely that $E(\ell)$ could have such an intricate structure, with at least three local extrema. We can strengthen this argument by assuming that the functional form of the matter field wavefunction $\psi$ must belong to a one-parametric family, parametrized by $\ell$. Then, on dimensional grounds, we must have $\psi(\mathbf{r})=\varphi(\mathbf{r}/\ell)/\ell^{d/2}$, where $\varphi(z)$ does not depend on $\ell$. From Eqs.~(\ref{E}-\ref{V}) it then follows that $E(\ell)=A/\ell^{p}-B/\ell^{d}$, where $A$ and $B$ are positive and do not depend on $\ell$. Then $E'(\ell)$ has a single zero, at $\ell_{0}=\left(\sfrac{dB}{pA}\right)^{1/(d-p)}$, for which \[E''(\ell_{0})=(p-d)pA\left(\frac{dB}{pA}\right)^{\frac{p+2}{p-d}}\,.\]
So, if $p<d$, $E''(\ell_{0})<0$, so $\ell_{0}$ is a global maximum, and there are no local or global minima. This seems the extent to which the heuristic argument can be pushed, and is one reason why it is useful to have the rigorous VK criterion.

The standard 1D nonlinear Schr\"{o}dinger equation ($p=2$, $d=1$) is known
to support solitons \cite{zakharov1972_62,solitons_book,strecker2002_150,khaykovich2002_1290} and indeed it satisfies the first inequality in Eq.~(\ref{VT_for_GPE}). To the contrary, in 3D, the localized structures are known
to collapse \cite{berde1998_259,donley2001_295}: and indeed, the combination $p=2$ and $d=3$ violates both inequalities in Eq.~(\ref{VT_for_GPE}). We will now suggest a way to boost the dispersion sharpness to $p=4$ using dispersion management in optical lattices, and in so doing
stabilize the 3D solitary waves.

\section{Realizing a quartic dispersion law in a shaken optical lattice}
Consider a single atom in a shaken 3D optical lattice:
\begin{align}
\hat{H} = \frac{\hat{p}^2}{2m}
+ W
\sum_{\alpha=x,\,y,\,z} \cos[\kappa(r_{\alpha}-\xi\cos\omega t)]
\,\,,
\label{hamiltonian}
\end{align}
where $m$ is the atomic mass, $2W$ the lattice depth, $\kappa=2\pi/\lat$ the lattice
wavevector, $\lat$ the lattice spacing, $\omega$ the shaking frequency,
$\xi$ the shaking amplitude (shaking is applied in the grand-diagonal
direction), and $r_{x,\,y,\,z} = x,\,y,\,z$.
We look for the solutions of the corresponding time-dependent
Schr\"{o}dinger equation using a separation-of-variables ansatz:
$\phi({\bm r},\,t) = \prod_{\alpha=x,\,y,\,z} \phi^{(\alpha)}(r_{\alpha},\,t)$.
The time-dependent Schr\"{o}dinger equations are identical for each of the
three factors. From now on, let us concentrate on the time evolution of
$\phi^{(x)}$. The results can be trivially recast for $\phi^{(y)}$ and $\phi^{(z)}$, and we will assemble them into a single 3D
solution in the end.

The time-evolution equation for $\phi^{(x)}$ is
\begin{align}
&
i \frac{\partial}{\partial t}\phi^{(x)} (x,\,t) = \hat{H}^{(x)} \phi^{(x)} (x,\,t)
\label{schrodinger_x}
\end{align}
with
\begin{align*}
\hat{H}^{(x)}
=
 \frac{\hat{p}_{x}^2}{2m}
 + W
\cos[\kappa(x-\xi\cos\omega t)]
\,\,,
%
\end{align*}
where $\hat{H}^{(x)}$ enters the full Hamiltonian through
$\hat{H} = \hat{H}^{(x)} + \hat{H}^{(y)} +\hat{H}^{(z)}$.
Assume that $\xi \ll \lat$ and Taylor expand $\hat{H}^{(x)}$ to the first power in $\xi$:
\begin{multline*}
%
\hat{H}^{(x)} \approx
\hat{H}^{(x)}_{0} + \hat{U}^{(x)}
=
\frac{\hat{p}_{x}^2}{2m}
+ W  \cos \kappa x
\\
\qquad\qquad
+
\kappa\xi  W  (\sin \kappa x) (\cos \omega t)
\,,
%
\end{multline*}
where the operators $\hat{H}^{(x)}_{0}=\frac{\hat{p}_{x}^2}{2m}
+ W  \cos \kappa x $ and
$\hat{U}^{(x)}= \kappa\xi W  (\sin \kappa x) (\cos \omega t)$ are regarded as an ``unperturbed Hamitonian''
and a ``perturbation,'' respectively. The corresponding time-dependent
Schr\"{o}dinger equation supports Floquet-type solutions,
\begin{align}
\phi^{(x)}({\bm r},\,t)
=
\left(\sum_{\nm =-\infty}^{+\infty} \phi^{(x)}_{\nm} ({\bm r}) e^{i \nm \omega t}\right)
e^{-i \mathcal{E} t/\hbar}\,,
\end{align}
where $\mathcal{E}$ is the Floquet quasi-energy.
The states
$|\chi^{(x)}(x)\rangle = \sum_{\nm=-\infty}^{+\infty} \phi^{(x)}_{\nm} (x) |\nm\rangle$
are the eigenstates of the Floquet Hamiltonian:
\begin{align}
&
\hat{\mathcal{H}}^{(x)} |\chi^{(x)}(x)\rangle = \mathcal{E}^{(x)}  |\chi^{(x)}(x)\rangle
\end{align}
with
\begin{multline}
\hat{\mathcal{H}}^{(x)} =
\frac{\hat{p}_{x}^2}{2m}
+ W (\cos \kappa x)
+ \sum_{\nm=-\infty}^{+\infty} \nm \hbar\omega \,|\nm \rangle\langle \nm|
\\
+
 \frac{1}{2} \kappa\xi W
(\sin \kappa x)
 \sum_{\nm=-\infty}^{+\infty} (|\nm+1\rangle \langle \nm| + |\nm\rangle \langle \nm+1| )
\,\,.
\label{hamiltonian_x_floquet}
\end{multline}
Here, the states $|\nm \rangle$ lie in a Floquet Hilbert space.

The  Floquet Hamiltonian in Eq.~(\ref{hamiltonian_x_floquet}) is also periodic in space. Following the Bloch theorem, we will be looking for the eigenstates of the Floquet Hamiltonian that have the form
\begin{align*}
|\chi^{(x),\, K_{x}}(x)\rangle =
\left(
\sum_{\ell=-\infty}^{+\infty}
|\chi^{(x),\, K_{x},\,{\ell}}\rangle \frac{1}{\sqrt{d}} \exp[i \ell\kappa x]
\right)
e^{i K_{x} x}
\,\,,
\end{align*}
where $K_{x} $ is the $x$-component of the Bloch vector.

\startsquarepar
Denote the eigenstates and the eigenvalues of the
Floquet Hamiltonian in Eq.~(\ref{hamiltonian_x_floquet}), labeled by indexes $\bar{\nm}=0,\,\pm 1,\,\pm 2,\,\ldots$ and $s=0,\,1,\,2,\,\ldots$,  as
$|\chi^{(x),\, K_{x},\,\bar{\nm},\,s}(x)\rangle$ and $\mathcal{E}^{(x),\,\bar{\nm},\,s}(K_{x})$ respectively. For zero shaking, the eigenstates and eigenenergies are related
to the eigenstates
\stopsquarepar
%
\noindent $\chi^{(x),\, K_{x},\,s}_{0}(x)$ and the eigenenergies $E^{(x),\,s}_{0}(K_{x})$ of the stationary lattice as:
\begin{align}
&
|\chi^{(x),\, K_{x},\,\bar{\nm},\,s}(x)\rangle \stackrel{\xi \to 0}{\longrightarrow} \chi^{(x),\, K_{x},\,s}_{0}(x) |\nm\!=\!\bar{\nm}\rangle
\label{Bloch-Floquet_chi}
\\
&
\mathcal{E}^{(x),\,\bar{\nm},\,s}(K_{x})  \stackrel{\xi\to 0}{\longrightarrow} E^{(x),\,s}_{0}(K_{x}) + \hbar\omega \bar{\nm}
\label{Bloch-Floquet_E}
\,\,.
\end{align}
\noindent For a shaking frequency close to the transition frequency between the ground and the first excited energy bands in the middle of the
Brillouin zone,
\begin{align}
\hbar\omega \approx E^{(x),\,s\!=\!1}_{0}(K_{x}\!=\!0) - E^{(x),\,s\!=\!0}_{0}(K_{x}\!=\!0)
\,\,,
\label{resonance}
\end{align}
these bands hybridize  \cite{gemelke2005_170404,lignier2007_220403,struck2011_996,struck2012_225304,parker2013_769}. Observe
that for a deep blue detuning, the bands will exchange roles and the parabolic dispersion law in the ground band will become an inverted parabola.
By continuity, for each shaking amplitude, there will exist a shaking frequency for which the quadratic term in the dispersion law vanishes, the quartic term becoming dominant. Below, we analyze a sample set of parameters where this phenomenon indeed happens. The resonance
condition in Eq.~(\ref{resonance}) allows us to use degenerate perturbation theory, truncating the Hilbert space to a 2D space spanned
by the states
$\chi^{(x),\, K_{x},\,s\!=\!0}_{0}(x) |\nm \!=\!0\rangle$ and $\chi^{(x),\, K_{x},\,s\!=\!1}_{0}(x) | \nm  \!=\!-1\rangle$, with the unperturbed energies
$E^{(x),\,s\!=\!0}_{0}(K_{x})$ and $E^{(x),\,s\!=\!1}_{0}(K_{x}) - \hbar\omega$, respectively. Our construction
of a 3D quartic dispersion curve is completed by the observation that the identical dispersion law will be generated in the other two
spatial directions, and that the resulting dispersion hypersurface is simply the sum of the three.

\section{An outline of a scheme for experimental realization}
\begin{figure}[H]
\centering\includegraphics[width=.45\textwidth]{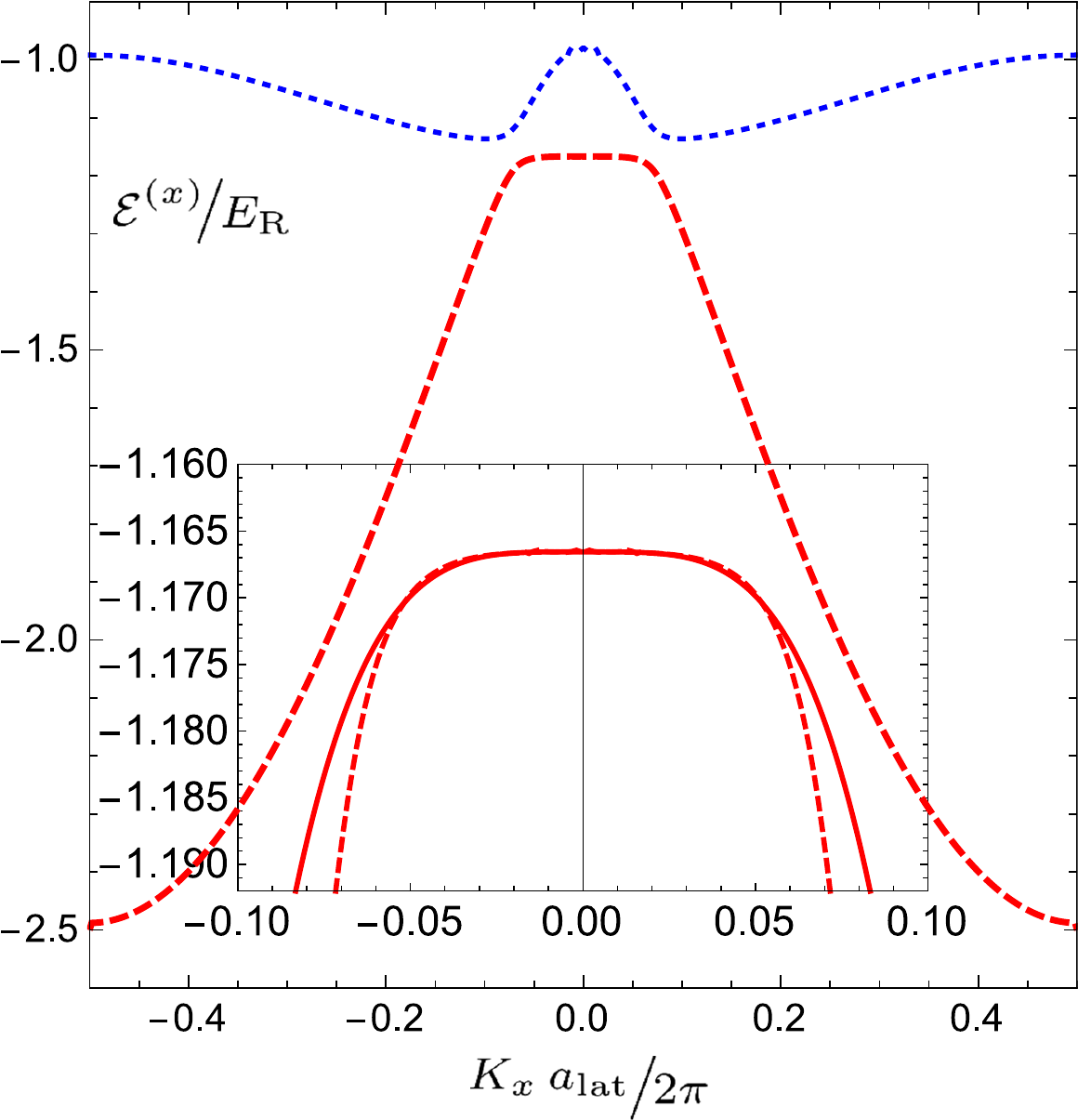}
\caption{(color online).
Floquet energies $\mathcal{E}^{(x),\,0,\,0}(K_{x})$ (red dashed lines) and $\mathcal{E}^{(x),\,-1,\,1}(K_{x})$
(blue dotted line), the adiabatic descendants of the ground and the first excited bands, respectively,
as functions of the Bloch vector $K_{x}$. The parameters used correspond
to a system of $^{133}$Cs atoms in an optical lattice with depth $2 W = 6.8\, E_{\text{R}}$ and
spacing $\lat=532\,\mbox{nm}$, shaken in the grand-diagonal direction
with frequency $\omega = 4.75\,E_{\mbox{\scriptsize R}}/\hbar$ and amplitude  $\sqrt{3} \times 0.0056\, d$. The curves are calculated using
degenerate perturbation theory applied to the Floquet Hamiltonian. For our set of parameters, quadratic terms in the ground band dispersion
are canceled, and the formerly sub-leading inverted quartic dispersion (the red solid line in the inset) becomes dominant. Dispersion laws in the two remaining directions, $y$ and $z$, are identical to the one along $x$. Inset: a magnified view of the peak of the Floquet energy $\mathcal{E}^{(x),\,0,\,0}(K_{x})$ (red dashed line), showing that it is very close to the pure---formerly sub-leading---inverted quartic dispersion (the red solid line). The axes labels of the main plot are also the axes labels for the plot in the inset.
}
\label{f:dispersion}
\end{figure}

Let us now discuss a sample experimental realization of the above scheme.
Consider a set of parameters inspired by
the scheme used to induce a long-range ferromagnetic order in an atomic gas \cite{parker2013_769},
as follows. $N=10^{3}$ $^{133}$Cs atoms are loaded in a 3D optical lattice with
lattice spacing $\lat=532\,\mbox{nm}$  and depth $2 W = 6.8\, E_{\mbox{\scriptsize R}}$, where $E_{\mbox{\scriptsize R}} = \hbar^2 k^2/2m$ is the
recoil energy, and $k=\kappa/2=\pi/\lat$ is the wavevector of the lattice light. The interactions are repulsive and some trapping will
also be required at this stage.
We are going to postpone the choice of the scattering length till later.
Initially, there is no shaking. Then a low-frequency shaking starts, with the
amplitude $\sqrt{3}\,\xi$, in the grand diagonal direction, where $\xi = 0.0056\, \lat$  is the shaking amplitude along each of the three 1D-lattice direction.
Subsequently, the shaking frequency is slowly ramped to $\omega = 4.75\,E_{\mbox{\scriptsize R}}/\hbar$, a frequency slightly to the red from the
interband spacing in the center of the Brillouin zone. For the same frequency, the edges of the zone undergo a complete adiabatic population inversion during the ramp.
The final shaking frequency is tuned in such a way that
the quadratic terms in the dispersion law of the---heavily hybridized---ground band are canceled: the remaining curve is
the 3D \textit{inverted} quartic parabola depicted in Fig.~\ref{f:dispersion}.
Following the analogy with the 1D gap solitons \cite{eiermann2004_230401}, we concentrate on the time-evolution
of the complex conjugate of the wavefunction. Its time evolution is governed by a non-inverted quartic dispersion law:
\begin{align}
i \hbar \frac{\partial \tilde{\Psi}}{\partial t}
&
=
\tilde{\eta}\, \hbar^4 \left(
\frac{\partial^4 \tilde{\Psi}}{\partial x^4} + \frac{\partial^4 \tilde{\Psi}}{\partial y^4}  + \frac{\partial^4 \tilde{\Psi}}{\partial z^4}\right)
+ \tilde{g} |\tilde{\Psi}|^2\tilde{\Psi}
\label{time-dependent_GPE_2}
\\
\tilde{\eta}>0,
&
\;\tilde{g}<0\,,
\notag
\end{align}
where $\tilde{\Psi}\equiv\Psi^{*}$, $\tilde{\eta} \equiv - \eta$, and $\tilde{g}\equiv -g$. Accordingly, the stability of the solitary wave will require repulsive interactions; as an example, we set the scattering
length to $a_{\mbox{\scriptsize S}} = 2.4\, a_{\mbox{\scriptsize Bohr}}$.
At this point,
trapping can be removed, following a numerically optimized schedule, and the resulting atomic cloud should be entirely self-supporting.
The dispersion constant  entering the time-dependent wave equation in Eq.~(\ref{time-dependent_GPE_2})  will have the value
\begin{align*}
\tilde{\eta}
\equiv  - 4! \, (d^4 \mathcal{E}^{(\alpha),\,0,\,0}(K_{\alpha})/d K_{\alpha}^4)_{K_{\alpha}=0}/\hbar^4
\,\,,
\end{align*}
while the coupling constant in Eq.~(\ref{time-dependent_GPE_2}) will be given by $\tilde{g} \equiv -\zeta^{3}\,\times
(4\pi\hbar^2 a_{\mbox{\scriptsize S}}/m)$, where
\[
\zeta = \lat \int_{-\lat/2}^{+\lat/2}\!dr_{\alpha} \,
|\langle \nm\!=\!0 | \chi^{(\alpha),\, 0,\,0,\,0}(r_{\alpha})\rangle|^4 \,;
\]
here $\alpha$ is any of the three Cartesian directions, $x$, $y$, or $z$.
Using our chosen set of parameters, this works out to $\tilde{\eta}  = 0.3 \, E_{\mbox{\scriptsize R}} (\lat/\hbar)^4$
and $\tilde{g} = -0.6\, E_{\mbox{\scriptsize R}} \lat^3/N$  (see Eqs.~(\ref{Bloch-Floquet_chi})
and (\ref{Bloch-Floquet_E}) for notation).
For the above set of parameters, a Gaussian variational
estimate predicts the wavevector distribution diameter of
$\Delta K \equiv 2 (\langle K^4 \rangle)^{1/4} \approx 0.018 (2\pi/\lat)$; 
for this degree of momentum localization, the solitary wave 
\noindent would occupy
only  $\sim 1/20$th of the Brillouin zone in each of the three directions, and thus it is expected to be fully mobile. (Above, we used the exact result in Eq.~(\ref{E_vs_T_virial})
to relate the mean fourth power of the quasi-momentum to the total energy.) This set of parameters also makes the interatomic dipole-dipole interaction small: its relative strength compared to the contact interaction is $\epsilon_{dd} = .25$ (see p.~70 of Ref.~\cite{GustavssonPhD2008}).

\section{Conclusion and outlook}
We have shown that a 3D Bose-Einstein condensate with a quartic dispersion law
satisfies the VK criterion for the stability of localized
structures in a dispersive non-relativistic self-focussing medium, and that such a dispersion law can be realized in shaken
optical lattices \cite{gemelke2005_170404,lignier2007_220403,struck2011_996,struck2012_225304,parker2013_769}.
Experimentally, as in the realization of band-gap
solitons  \cite{eiermann2004_230401}, our scheme will involve an inverse quartic dispersion and repulsive condensates.

We see rotation sensing as the primary area of application of 3D solitary waves. Interferometry with self-attracting waves was a success from the
onset \cite{mcdonald2014_013002} (see also \cite{scott2008_100402,nguyen2014_918,marchant2013_1865,kasevich2012_NASA}), showing a manyfold increase in
the fringe visibility as compared to free waves. Further improvements in sensitivity are expected for slow splitting of solitary waves on barriers \cite{hansen2012_12101681} where the appearance of macroscopic quantum superpositions is predicted \cite{weiss2009_010403,streltsov2009_043616}.
But while linear solitary wave interferometers are ideally suited for acceleration sensing, Sagnac measurements require a multi-dimensional geometry. There are waveguide-based proposals \cite{Helm2015_134101}; their principal drawback---in addition to the fact that the loading of atoms into the guide is difficult, and that the requirements for the stability of the guiding fields are very strict---is the inflexibility of the interferometric scheme. In developing an interferometric scheme, one starts with a seed idea and then builds upon it by successive modifications--usually strengthening the beamsplitters in some way---that result in increases in the interferometric area (space-time for the gravitometers/accelerometers, and space-space for the rotation sensors). For waveguide-based schemes, this requires rebuilding the waveguide each time, with all the accompanying optimization and stabilization issues. In contrast, schemes based on free-space propagation allow for easy implementation of improvements. The present work paves the way for such schemes, by proposing a method to obtain solitary waves that propagate in free space.

%
%
\section*{Acknowledgements}
We thank David Campbell for his remarks.
This work was supported by grants from the US National Science Foundation ({\it PHY-1402249}) and
the US Office of Naval Research ({\it N00014-12-1-0400}).


\bibliographystyle{elsarticle-num}
\bibliography{Olshanii_p4_matter_waves_02}

\begin{thebibliography}{10}
\expandafter\ifx\csname url\endcsname\relax
  \def\url#1{\texttt{#1}}\fi
\expandafter\ifx\csname urlprefix\endcsname\relax\def\urlprefix{URL }\fi
\expandafter\ifx\csname href\endcsname\relax
  \def\href#1#2{#2} \def\path#1{#1}\fi

\bibitem{ablowitz1981}
M.~J. Ablowitz, H.~Segur, Solitons and the Inverse Scattering Transform, SIAM,
  Philadelphia, 1981.
\newblock \href {http://dx.doi.org/10.1137/1.9781611970883}
  {\path{doi:10.1137/1.9781611970883}}.

\bibitem{strecker2002_150}
K.~E. Strecker, G.~B. Partridge, A.~G. Truscott, R.~G. Hulet, Formation and
  propagation of matter wave soliton trains, Nature 417 (2002) 150.
\newblock \href {http://dx.doi.org/10.1038/nature747}
  {\path{doi:10.1038/nature747}}.

\bibitem{khaykovich2002_1290}
L.~Khaykovich, F.~Schreck, G.~Ferrari, T.~Bourdel, J.~Cubizolles, L.~D. Carr,
  Y.~Castin, C.~Salomon, Formation of a matter-wave bright soliton, Science 296
  (2002) 1290.
\newblock \href {http://dx.doi.org/10.1126/science.1071021}
  {\path{doi:10.1126/science.1071021}}.

\bibitem{ahufinger2004_053604}
V.~Ahufinger, A.~Sanpera, P.~Pedri, L.~Santos, M.~Lewenstein, Creation and
  mobility of discrete solitons in {B}ose-{E}instein condensates, Phys. Rev. A
  69 (2004) 053604.
\newblock \href {http://dx.doi.org/10.1103/PhysRevA.69.053604}
  {\path{doi:10.1103/PhysRevA.69.053604}}.

\bibitem{alexander2006_040401}
T.~J. Alexander, E.~A. Ostrovskaya, Y.~S. Kivshar, Self-trapped nonlinear
  matter waves in periodic potentials, Phys. Rev. Lett. 96 (2006) 040401.
\newblock \href {http://dx.doi.org/10.1103/PhysRevLett.96.040401}
  {\path{doi:10.1103/PhysRevLett.96.040401}}.

\bibitem{saito2004_053610}
H.~Saito, M.~Ueda, {B}ose-{E}instein droplet in free space, Phys. Rev. A 70
  (2004) 053610.
\newblock \href {http://dx.doi.org/10.1103/PhysRevA.70.053610}
  {\path{doi:10.1103/PhysRevA.70.053610}}.

\bibitem{mihalache2005_021601}
D.~Mihalache, D.~Mazilu, F.~Lederer, B.~A. Malomed, L.-C. Crasovan, Y.~V.
  Kartashov, L.~Torner, Stable three-dimensional solitons in attractive
  {B}ose-{E}instein condensates loaded in an optical lattice, Phys. Rev. A 72
  (2005) 021601.
\newblock \href {http://dx.doi.org/10.1103/PhysRevA.72.021601}
  {\path{doi:10.1103/PhysRevA.72.021601}}.

\bibitem{borovkova2011_035602}
O.~V. Borovkova, Y.~V. Kartashov, L.~Torner, B.~A. Malomed, Bright solitons
  from defocusing nonlinearities, Phys. Rev. E 84 (2011) 035602.
\newblock \href {http://dx.doi.org/10.1103/PhysRevE.84.035602}
  {\path{doi:10.1103/PhysRevE.84.035602}}.

\bibitem{vakhitov1973_1020}
N.~Vakhitov, A.~Kolokolov, Stationary solutions of the wave equation in the
  medium with nonlinearity saturation, Izv. Vuz. Radiofiz. 16 (1973) 1020,
  [Sov. J. Radiophys. Quantum Electron. {\bf 16}, 783 (1973)].

\bibitem{book_landau_mechanics}
L.~Landau, E.~Lifshitz, Mechanics: Volume 1 (Course of Theoretical Physics
  Series), Butterworth-Heinemann, Amsterdam, 1976.

\bibitem{zakharov1972_62}
V.~E. Zakharov, A.~B. Shabat, Exact theory of two-dimensional self-focusing and
  one-dimensional self-modulation of waves in nonlinear media, Soviet Physics
  JETP 34 (1972) 62.

\bibitem{solitons_book}
P.~G. Drazin, R.~S. Johnson, Solitons: an Introduction, Cambridge University
  Press, New York, 1989.

\bibitem{berde1998_259}
L.~Berg\'{e}, Wave collapse in physics: principles and applications to light
  and plasma waves, Physics Reports 303 (1998) 256.
\newblock \href {http://dx.doi.org/10.1016/S0370-1573(97)00092-6}
  {\path{doi:10.1016/S0370-1573(97)00092-6}}.

\bibitem{donley2001_295}
E.~A. Donley, N.~R. Claussen, S.~L. Cornish, J.~L. Roberts, E.~A. Cornell,
  C.~E. Wieman, Dynamics of collapsing and exploding {B}ose-{E}instein
  condensates, Nature 412 (2001) 295.

\bibitem{gemelke2005_170404}
N.~Gemelke, E.~Sarajlic, Y.~Bidel, S.~Hong, S.~Chu, Parametric amplification of
  matter waves in periodically translated optical lattices, Phys. Rev. Lett. 95
  (2005) 170404.
\newblock \href {http://dx.doi.org/10.1103/PhysRevLett.95.170404}
  {\path{doi:10.1103/PhysRevLett.95.170404}}.

\bibitem{lignier2007_220403}
H.~Lignier, C.~Sias, D.~Ciampini, Y.~Singh, A.~Zenesini, O.~Morsch,
  E.~Arimondo, Dynamical control of matter-wave tunneling in periodic
  potentials, Phys. Rev. Lett. 99 (2007) 220403.
\newblock \href {http://dx.doi.org/10.1103/PhysRevLett.99.220403}
  {\path{doi:10.1103/PhysRevLett.99.220403}}.

\bibitem{struck2011_996}
J.~Struck, C.~\"Olschl\"ager, R.~Le~Targat, P.~Soltan-Panahi, A.~Eckardt,
  M.~Lewenstein, P.~Windpassinger, K.~Sengstock, Quantum simulation of
  frustrated classical magnetism in triangular optical lattices, Science 333
  (2011) 996.
\newblock \href {http://dx.doi.org/10.1126/science.1207239}
  {\path{doi:10.1126/science.1207239}}.

\bibitem{struck2012_225304}
J.~Struck, C.~\"Olschl\"ager, M.~Weinberg, P.~Hauke, J.~Simonet, A.~Eckardt,
  M.~Lewenstein, K.~Sengstock, P.~Windpassinger, Tunable gauge potential for
  neutral and spinless particles in driven optical lattices, Phys. Rev. Lett.
  108 (2012) 225304.
\newblock \href {http://dx.doi.org/10.1103/PhysRevLett.108.225304}
  {\path{doi:10.1103/PhysRevLett.108.225304}}.

\bibitem{parker2013_769}
C.~V. Parker, L.-C. Ha, C.~Chin, Direct observation of effective ferromagnetic
  domains of cold atoms in a shaken optical lattice, Nat. Phys. 9 (2013) 769.
\newblock \href {http://dx.doi.org/10.1038/nphys2789}
  {\path{doi:10.1038/nphys2789}}.

\bibitem{eiermann2004_230401}
B.~Eiermann, T.~Anker, M.~Albiez, M.~Taglieber, P.~Treutlein, K.-P. Marzlin,
  M.~K. Oberthaler, Bright {B}ose-{E}instein gap solitons of atoms with
  repulsive interaction, Phys. Rev. Lett. 92 (2004) 230401.
\newblock \href {http://dx.doi.org/10.1103/PhysRevLett.92.230401}
  {\path{doi:10.1103/PhysRevLett.92.230401}}.

\bibitem{GustavssonPhD2008}
M.~Gustavsson,
  \href{http://www.ultracold.at/theses/thesis_mattias_gustavsson/Thesis_Mattias_Gustavsson.pdf}{A
  quantum gas with tunable interactions in an optical lattice}, Ph.D. thesis,
  University of Innsbruck (2008).
\newline\urlprefix\url{http://www.ultracold.at/theses/thesis_mattias_gustavsson/Thesis_Mattias_Gustavsson.pdf}

\bibitem{mcdonald2014_013002}
G.~D. McDonald, C.~C.~N. Kuhn, K.~S. Hardman, S.~Bennetts, P.~J. Everitt, P.~A.
  Altin, J.~E. Debs, J.~D. Close, N.~P. Robins, Bright solitonic matter-wave
  interferometer, Phys. Rev. Lett. 113 (2014) 013002.
\newblock \href {http://dx.doi.org/10.1103/PhysRevLett.113.013002}
  {\path{doi:10.1103/PhysRevLett.113.013002}}.

\bibitem{scott2008_100402}
R.~G. Scott, T.~E. Judd, T.~M. Fromhold, Exploiting soliton decay and phase
  fluctuations in atom chip interferometry of {B}ose-{E}instein condensates,
  Phys. Rev. Lett. 100 (2008) 100402.
\newblock \href {http://dx.doi.org/10.1103/PhysRevLett.100.100402}
  {\path{doi:10.1103/PhysRevLett.100.100402}}.

\bibitem{nguyen2014_918}
H.~Nguyen, P.~Dyke, D.~Luo, B.~A. Malomed, R.~G. Hulet, Collisions of
  matter-wave solitons, Nat. Phys. 10 (2014) 918.
\newblock \href {http://dx.doi.org/10.1038/nphys3135}
  {\path{doi:10.1038/nphys3135}}.

\bibitem{marchant2013_1865}
A.~L. Marchant, T.~P. Billam, T.~P. Wiles, M.~M.~H. Yu, S.~A. Gardiner, S.~L.
  Cornish, Controlled formation and reflection of a bright solitary
  matter-wave, Nat. Commun. 4 (2013) 1865.
\newblock \href {http://dx.doi.org/10.1038/ncomms2893}
  {\path{doi:10.1038/ncomms2893}}.

\bibitem{kasevich2012_NASA}
M.~Kasevich, Atom systems and {B}ose {E}instein condensates for metrology and
  navigation, first NASA Quantum Future Technologies Conference (2012).

\bibitem{hansen2012_12101681}
S.~D. Hansen, N.~Nygaard, K.~M. lmer, Scattering of matter wave solitons on
  localized potentials, Preprint at arXiv:1210.1681 (2012).

\bibitem{weiss2009_010403}
C.~Weiss, Y.~Castin, Creation and detection of a mesoscopic gas in a nonlocal
  quantum superposition, Phys. Rev. Lett. 102 (2009) 010403.
\newblock \href {http://dx.doi.org/10.1103/PhysRevLett.102.010403}
  {\path{doi:10.1103/PhysRevLett.102.010403}}.

\bibitem{streltsov2009_043616}
A.~I. Streltsov, O.~E. Alon, L.~S. Cederbaum, Scattering of an attractive
  {B}ose-{E}instein condensate from a barrier: {F}ormation of quantum
  superposition states, Phys. Rev. A 80 (2009) 043616.
\newblock \href {http://dx.doi.org/10.1103/PhysRevA.80.043616}
  {\path{doi:10.1103/PhysRevA.80.043616}}.

\bibitem{Helm2015_134101}
J.~L. Helm, S.~L. Cornish, S.~A. Gardiner, Sagnac interferometry using bright
  matter-wave solitons, Phys. Rev. Lett. 114 (2015) 134101.
\newblock \href {http://dx.doi.org/10.1103/PhysRevLett.114.134101}
  {\path{doi:10.1103/PhysRevLett.114.134101}}.

\end{thebibliography}

\end{document}